\documentclass[
twocolumn, 
superscriptaddress, 
nofootinbib, 
preprintnumbers, 
prl 
]{revtex4-2} 

\usepackage[centering,hmargin=2.5cm,vmargin=2.5cm]{geometry}
\usepackage[colorlinks=true,citecolor=blue,linkcolor=blue]{hyperref}
\usepackage[normalem]{ulem}
\usepackage{amsmath,amssymb}
\usepackage{mathtools}
\usepackage{epsfig}
\usepackage{graphicx}               
\usepackage{url}
\usepackage{color}
\usepackage{multirow}
\usepackage{placeins}
\usepackage[dvipsnames]{xcolor}
\usepackage{epstopdf}
\usepackage{siunitx}
\usepackage{enumitem}
\usepackage{cleveref}
\usepackage{lipsum}
\usepackage{gensymb}
\usepackage{xspace}
\usepackage{titlesec}
 
\allowdisplaybreaks

\newcommand{\bra}[1]{\ensuremath{\langle #1 |}}   
\newcommand{\ket}[1]{\ensuremath{| #1 \rangle}}   
\newcommand{\ev}[1]{\ensuremath{\left\langle #1 %
        \right\rangle}} 

\newcommand{\upd}{\mathrm d}                                          

\newcommand{\f}{\ensuremath{f_\chi}\xspace}

\titleformat*{\section}{\centering\bfseries\scshape}
\titlelabel{\thetitle\quad}
\titleformat*{\paragraph}{\bfseries}
\titlespacing*{\paragraph}{0pt}{3.25ex plus 1ex minus .2ex}{1em}


\begin{document}

\title{Primordial Black Holes from First-Order Cosmological Phase Transitions}

\author{Michael J.\ Baker}
\email{michael.baker@unimelb.edu.au}
\affiliation{ARC Centre of Excellence for Dark Matter Particle Physics, 
School of Physics, The University of Melbourne, Victoria 3010, Australia}

\author{Moritz Breitbach}
\email{breitbach@uni-mainz.de}
\affiliation{PRISMA Cluster of Excellence \& Mainz Institute for
    Theoretical Physics, \\
    Johannes Gutenberg University, Staudingerweg 7, 55099
    Mainz, Germany}

\author{Joachim Kopp}
\email{jkopp@cern.ch}
\affiliation{PRISMA Cluster of Excellence \& Mainz Institute for
    Theoretical Physics, \\
    Johannes Gutenberg University, Staudingerweg 7, 55099
    Mainz, Germany}
\affiliation{Theoretical Physics Department, CERN,
    Esplanade des Particules, 1211 Geneva 23, Switzerland}

\author{Lukas Mittnacht}
\email{lmittna@uni-mainz.de}
\affiliation{PRISMA Cluster of Excellence \& Mainz Institute for
    Theoretical Physics, \\
    Johannes Gutenberg University, Staudingerweg 7, 55099
    Mainz, Germany}

\date{\today}

\preprint{CERN-TH-2021-079}
\preprint{MITP-21-023}


\begin{abstract}
    \noindent
    We discuss the possibility of forming primordial black holes during a first-order phase transition in the early Universe.  As is well known, such a phase transition proceeds through the formation of true-vacuum bubbles in a Universe that is still in a false vacuum. When there is a particle species whose mass increases significantly during the phase transition, transmission of the corresponding particles through the advancing bubble walls is suppressed. Consequently, an overdensity can build up in front of the walls and become sufficiently large to trigger primordial black hole formation. We track this process quantitatively by solving a Boltzmann equation, and we delineate the phase transition properties required for our mechanism to yield an appreciable abundance of primordial black holes.
\end{abstract}

\maketitle

\textbf{Introduction.}
%
Black holes are among the most fascinating objects in the Universe, inspiring not only scientists, but also science fiction authors, filmmakers and their audiences. Since Schwarzschild~\cite{1916AbhKP1916..189S} and Droste~\cite{1917KNAB...19..197D} first provided exact solutions to the Einstein equations, which predict the existence of black holes, the field has come a long way. In 2016 the LIGO and Virgo collaborations announced the first observation of gravitational waves from a merger of stellar-mass black holes~\cite{Abbott:2016blz}, followed in 2019 by the spectacular first direct image of a supermassive black hole by the Event Horizon Telescope~\cite{Akiyama:2019cqa}. 

While we know that black holes can form when old, massive stars collapse after running out of fuel, it is not clear whether this creation mechanism can account for the whole population of observed black holes. There is therefore a keen interest in \emph{primordial black holes} (PBHs), which are created in the hot and dense early Universe fractions of a second after the Big Bang.  For recent reviews on PBHs, see refs.~\cite{
Carr:2016drx,                  
Sasaki:2018dmp,                
Green:2020jor,                 
Carr:2020xqk,                  
Carr:2020gox,
Villanueva-Domingo:2021spv}.   
Depending on the time at which they form, PBHs can have almost any mass, and could solve a number of problems in cosmology: most importantly, they could constitute all or part of the dark matter (DM)~\cite{%
Carr:2016drx,                  
Green:2020jor,                 
Carr:2020xqk,                  
Villanueva-Domingo:2021spv},   
they could create other DM particles when they evaporate~\cite{
Green:1999yh,         
Khlopov:2004tn,       
Fujita:2014hha,       
Allahverdi:2017sks,
Lennon:2017tqq,       
Morrison:2018xla,     
Hooper:2019gtx,       
Masina:2020xhk,       
Baldes:2020nuv,       
Gondolo:2020uqv,      
Bernal:2020kse,       
Bernal:2020bjf,       
Bernal:2020bjf},      
change the expansion history of the Universe~\cite{%
Hooper:2019gtx,       
Chaudhuri:2020wjo},   
remove unwanted monopoles or domain walls from the Universe~\cite{
Stojkovic:2004hz, 
Stojkovic:2005zh},
or provide seeds for the supermassive black holes observed at the centre of galaxies~\cite{
Bean:2002kx}
or for large scale structure formation~\cite{Hoyle:1966,Ryan:1972,Carr:1984,Afshordi:2003zb,Carr:2020gox}.
There are several possible production mechanisms of PBHs: the most widely studied is collapse of density perturbations generated during inflation~\cite{
Carr:1975qj,
Ivanov:1994pa,
Garcia-Bellido:1996mdl,
Silk:1986vc,
Kawasaki:1997ju,
Yokoyama:1995ex,
Pi:2017gih}, 
while the collapse of topological defects~\cite{
Hawking:1987bn,
Polnarev:1988dh,
MacGibbon:1997pu,
Rubin:2000dq,
Rubin:2001yw,
Kusenko:2020pcg,
Ashoorioon:2020hln,
Brandenberger:2021zvn}, 
the dynamics of scalar condensates~\cite{
Cotner:2016cvr,
Cotner:2019ykd},
or collisions of bubble walls during a first-order phase transition~\cite{
Crawford:1982yz,
Kodama:1982sf,
  Moss:1994pi,          
  Freivogel:2007fx,     
  Hawking:1982ga,       
  Johnson:2011wt}       
are viable alternatives. 

We here present a new mechanism of PBH production during a first-order cosmological phase transition. While previous papers on this topic have only considered the energy density stored in the bubble wall, we will focus on a population of particles that interact with the bubble wall.  The mass of these particles may increase significantly during phase transitions due to either confinement or a Higgs mechanism.  If this is the case, energy conservation means that all but the highest energy particles will be reflected and remain in the false vacuum.  This scenario has been previously considered for other reasons, but black hole formation has not been shown.  For example, confining phase transitions have been shown to produce quark nugget DM \cite{Witten:1984rs, Zhitnitsky:2002qa, Oaknin:2003uv, Lawson:2013bya, Krylov:2013qe, Huang:2017kzu, Bai:2018dxf, Asadi:2021pwo}, where Standard Model (SM) or dark quarks are trapped in a colour-superconducting phase, unable to transition to the standard QCD vacuum.  The dynamics of reflected particles during a phase transition realising a Higgs mechanism has recently been studied in ref.~\cite{Hong:2020est}, where the focus was on the formation of Fermi-ball DM. In this scenario, small pockets of the false vacuum are supported by dark particles that have insufficient energy to transition to the true vacuum.  These false vacua are stable but not dense enough to collapse into black holes. In this work, we determine the conditions under which the density of reflected particles could lead to the formation of PBHs. The mass and abundance of the PBHs depend on the temperature at which the phase transition occurs and the probability that a black hole will form in a given volume.  We here present the general mechanism and describe quantitative results for a toy model.  Technical details of the calculation appear in ref.~\cite{Baker:2021sno}.

\begin{figure}
    \centering
    \includegraphics{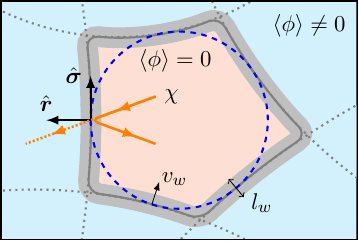}
    \caption{
      A cartoon picture of the late stage of a first-order cosmological phase transition: regions of true vacuum (blue) are expanding with speed $v_w$ and coalescing, leaving an approximately spherical bubble of false vacuum (light red).  Very high momentum $\chi$ particles can pass through the bubble wall into the true vacuum and gain a large mass, while lower momentum $\chi$ particles are reflected due to energy conservation.  The build-up of $\chi$ particles creates a density perturbation which may lead to PBH formation.  The local coordinate system is also shown, along with the bubble wall thickness, $l_w$.}
    \label{fig:illustration}
\end{figure}

\textbf{General Mechanism.}
%
We envision a scenario where there are two new fields, $\chi$ and $\phi$.  The field $\phi$ is a scalar and undergoes a first-order phase transition while in thermal contact with the SM.  At high temperatures $\phi$ does not have a vacuum expectation value (vev), $\ev{\phi} \equiv \bra{0}\phi\ket{0} = 0$, but at the nucleation temperature, $T_n$, $\phi$ develops a vev, $\ev{\phi} \neq 0$.  In first-order phase transitions bubble walls, where $\ev{\phi}$ transitions smoothly from 0 to its final value, move through the Universe, leaving behind the new $\ev{\phi} \neq 0$ phase.  The field $\chi$ interacts with $\phi$ in such a way that its mass significantly increases when $\phi$ obtains its vev. In the simplest case, this is achieved through a Yukawa coupling of the form $\mathcal{L}_Y = y_\chi \phi \overline\chi \chi$.  When a $\chi$ particle collides with the bubble wall, it can only pass through if it has enough kinetic energy to overcome the mass difference (due to energy conservation). Otherwise, the bubble walls will sweep up $\chi$ particles and increase their density in the $\ev{\phi} = 0$ region.  The resulting density perturbations can then lead to PBH formation.  The scenario is depicted schematically in \cref{fig:illustration}.  While most treatments of first-order phase transitions envision an expanding bubble with $\ev{\phi} \neq 0$ inside, we consider a \emph{shrinking} bubble with $\ev{\phi} = 0$ inside. The latter situation is more suitable for describing the late stages of the phase transition, when the density of particles swept along becomes highest and black hole formation is most likely.

We first perform a simplified analysis, to demonstrate the plausibility of the mechanism.  We take the bubble to have an initial radius $r_w^0 \sim r_H$ where $r_H \equiv 1/H$ (with $H$ the Hubble parameter) is the radius of a Hubble volume, i.e., the largest causally connected volume at a given time in the evolution of the Universe.

We will take $\chi$ to be relativistic inside the bubble and assume for the moment that no $\chi$ particles can penetrate the bubble wall. We assume $\chi$ begins in equilibrium, but as $\chi$ particles would actually have been swept up from the beginning of the phase transition, this assumption underestimates the achievable overdensity and is therefore conservative.  The density of the $\chi$ particles will increase as the bubble shrinks, with an energy density given by $\rho_\chi = [r_w^0/r_w(t)]^4 \rho_\chi^\text{eq}$, where $\rho_\chi^\text{eq} = (7\pi^2/240) g_\chi T_n^4$ is the equilibrium energy density of $\chi$ and $r_w(t)$ is the bubble radius at time $t$. The factor $g_\chi = 4$ accounts for the number of degrees of freedom of $\chi$ and its anti-particle $\bar{\chi}$.  This scaling can be understood from the increase in number density $\propto [r_w^0/r_w(t)]^3$ due to the decreasing bubble volume, combined with the energy gain $\propto r_w^0/r_w(t)$ of each $\chi$ particle due to reflections off a bubble wall moving at non-relativistic speeds.  On each reflection, the particle's energy increases by $\upd E = 2 v_w E$. The time between such collisions is $\upd t \approx 2 r_w(t) = 2 (r_w^0 - v_w t)$. From these two relations, we can deduce that $E \propto r_w^0/r_w(t)$.  While this energy increase on reflection is ultimately sourced from the latent heat of the phase transition, it is independent of the latent heat (and derived quantities such as $v_w$) when expressed in terms of the wall radius, as long as the wall moves at a constant, non-relativistic speed. We will assume that this is the case; more justification for this assumption is given below as well as in ref.~\cite{Baker:2021sno}.

A black hole forms when the bubble is smaller than the Schwarzschild radius of the total enclosed energy,
\begin{align}
    r_w(t) < 
    r_s \equiv 2 G E_\text{tot}^{(<r_w)} \,,
\end{align}
where $G$ is the gravitational constant and $E_\text{tot}^{(<r_w)}$ is the total energy inside a bubble of radius $r_w(t)$.  To be conservative, we will only include the energy of the $\chi$ particles and not that of, for instance, the wall itself.

On substituting in $\rho_\chi$ and using the Friedmann equation $H^2 = \frac{8}{90} \pi^3 G \, g_\star T_n^4$, with $g_\star$ the effective number of relativistic degrees of freedom in the thermal bath, we obtain
\begin{align}
\label{eq:schwarzschild-criteria}
    \frac{r_w(t)}{r_H} & < \sqrt{\frac{7}{8} \frac{g_\chi}{g_\star}}
                            \left(\frac{r_w^0}{r_H}\right)^{\!2} \,.
\end{align}
We see that the Schwarzschild criteria depends on temperature only through $g_\star(T_n)$, and is otherwise independent of any physical scale. For $T_n \gtrsim 100\,\text{GeV}$ (so $g_\star \sim 100$) and $r_w^0 \approx 1.5\,r_H$, the condition is satisfied when $r_w(t) \approx r_w^0/3.7$.  So if large $r_w^0 \sim r_H$ bubbles can capture the majority of the $\chi$ particles while shrinking by a factor of a few, then we would expect to form black holes.  

While this is encouraging, there are many physical effects not captured by this heuristic description: 
(1) Walls moving relativistically will impart a larger momentum boost on the $\chi$ particles. This effect increases the energy density of the $\chi$ particles, making it easier to satisfy the Schwarzschild criteria.  
(2) Sufficiently energetic $\chi$ particles will be able to traverse the wall, reducing the overdensity. 
(3) As $\rho_\chi$ grows, the gravitational force will further reinforce the overdensity.
(4) The $\chi$ particles may annihilate via $\chi\bar\chi \to \phi$ and $\chi\bar\chi \to \phi\phi$, and 
(5) they may scatter off each other and off $\phi$ particles, leading to momentum redistribution. 
(6) The wall velocity and width may vary as the bubble shrinks.
(7) Hubble expansion delays the shrinkage of the bubble, and red-shifts the energy density inside.

In the following we will discuss numerical simulations that take into account the first four of these effects.  The fifth one -- momentum redistribution -- is expected to be relatively unimportant as the scattering rate of $\chi$ is of the same order as the $\chi\bar\chi$ annihilation rate, and as black hole formation can be successful only when annihilation is negligible. Although the mechanism is approximately independent of the wall width and velocity -- the sixth point -- the time-dependence of $v_w$ can be important in extreme cases: a runaway bubble wall would mean that essentially all $\chi$ particles can pass through the bubble wall, while a high $\chi$ density could stop the wall completely, halting the phase transition.  Avoiding both of these extremes is possible if the latent heat release is not too large and not too small. In addition, complete models will often feature friction mechanism such as particle production at the bubble wall that prevent $v_w$ from becoming too large, while not impeding the advancement of the wall at slow $v_w$. The latent heat release should also not be so large that the interior of the bubble is vacuum energy dominated.   We discuss these aspects in more detail in a companion paper, ref.~\cite{Baker:2021sno}.  Finally, we neglect Hubble expansion because we will find that, if black holes form, they do so after not much more than one Hubble time.  We discuss all of these effects in more detail in ref.~\cite{Baker:2021sno}.

\textbf{Numerical Results.}
%
As a basis for our numerical analyses, we introduce a Dirac fermion $\chi$ and a real scalar field $\phi$, whose dynamics control a first-order phase transition.  Both $\chi$ and $\phi$ are SM gauge singlets.  The Lagrangian of our toy model contains the terms
\begin{align}
  \mathcal L \supset - m_\chi^0 \overline{\chi}\chi - y_\chi \phi\overline{\chi}{\chi} - V(\phi) \,,
\end{align}
where $y_\chi$ is a real Yukawa coupling and $V(\phi)$ is the scalar potential. When $\phi$ obtains a vev, $\chi$ obtains a mass correction $m_\chi = m_\chi^0 + y_\chi\!\ev{\phi}$.  In this toy model we will assume that $m_\chi^0 \ll T_n \ll y_\chi\!\ev{\phi}$, so $\chi$ will be relativistic inside the bubble.  
While $\chi$ also obtains a thermal mass, this will be negligible compared to $T_n$.
We will not assume any particular form for the scalar potential but will assume that it leads to a first-order phase transition, which we will parameterise by a nucleation temperature, $T_n$, order parameter, $\ev{\phi}\!/T_n$, wall thickness, $l_w$, and wall velocity, $v_w$. We further assume that the temperature does not vary significantly during the phase transition and that $\phi$ stays in thermal contact with the SM bath.  We also find that the mechanism requires large order parameters, $\ev{\phi} \gg T_n$, so that $m_\chi \gg T_n$. This is can be achieved in renormalizable polynomial potentials with only very mild fine tuning. 

Starting from the initial conditions discussed above, we then track the phase space distribution function of $\chi$, $f_\chi$, by numerically solving the Boltzmann equation $\boldsymbol{\mathrm L}[\f]\!=\!\boldsymbol{\mathrm C}[\f]$, where the Liouville operator $\boldsymbol{\mathrm L}[\f]$ accounts for the evolution of $f_\chi$ in the absence of hard interactions. The collision term $\boldsymbol{\mathrm C}[\f]$ accounts for particle annihilation and creation via $\chi\bar\chi \leftrightarrow \phi$ and $\chi\bar\chi \leftrightarrow \phi\phi$.  Assuming spherical symmetry reduces the problem to one spatial dimension and two momentum dimensions, so that $f_\chi \equiv f_\chi(r, p_r, p_\sigma, t)$ only retains a dependence on the radial coordinate $r$, on the momenta in the radial and tangential directions, $p_r$ and $p_\sigma$, and on time $t$.  We then solve the resulting system of equations numerically, building on methods similar to those described in ref.~\cite{Baker:2019ndr}. We use the method of characteristics, which transforms the Boltzmann equation -- a partial differential equation -- into a system of ordinary differential equations, each of which corresponds to a particle's phase space trajectory.  Full details of the equations and the methods used to solve them can be found in ref.~\cite{Baker:2021sno}.

\begin{figure}
    \centering
    \includegraphics{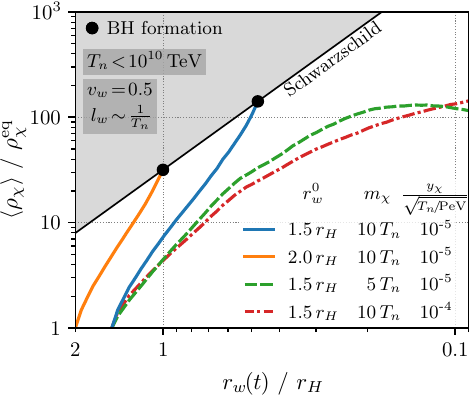}
    \caption{The simulated spatially-averaged energy density of $\chi+\bar\chi$, $\ev{\rho_\chi}$, inside a shrinking bubble as a function of time, parameterised by the bubble radius, for different parameter points. In the grey region the Schwarzschild criteria is satisfied, and the point of black hole formation is indicated by a black point.}
    \label{fig:result}
\end{figure}

In \cref{fig:result} we show the spatially averaged $\chi+\bar\chi$ energy density, $\ev{\rho_\chi}$, inside the bubble as a function of time, parameterised by the bubble radius, for four benchmark parameter choices.  We see that the blue curve with $r_w^0 = 1.5\,r_H$ satisfies the Schwarzschild criteria for black hole formation when it has shrunk by a factor of around three, as expected from \cref{eq:schwarzschild-criteria}.  The orange curve, characterised by a larger $r_w^0 = 2\,r_H$, satisfies the Schwarzschild criteria when it has shrunk to a radius of $\approx r_H$ (that is, by a factor of around two).  The average energy density, $\ev{\rho_\chi}$, increases approximately like a power-law $\propto r_w^{-4}(t)$, for the blue and orange curves. 

The dashed green and dot-dashed red curves show two parameter points which fail to lead to black hole formation.  For the parameter point shown in green, the mass gain ($m_\chi = 5 \, T_n$) is too small to prevent $\chi$ particles from leaking through the bubble wall.  In the red curve, the large Yukawa coupling leads to too many $\chi$ particles annihilating.  For mass gains larger, and Yukawa couplings smaller, than shown in the blue and orange curves, black hole formation is again successful.

Overall, we see from \cref{fig:result} that black hole formation is successful, as long as $y_\chi$ is small enough to avoid excessive $\chi$ annihilation, and $m_\chi / T_n$ is large enough to suppress $\chi$ transmission through the bubble wall. As $m_\chi \approx y_\chi\!\ev{\phi}$, this underlines the need for very large $\ev{\phi}\!/ T_n$.

\textbf{The Resulting Black Hole Population.}
%
The mass of a PBH that forms during radiation domination is $m_\text{BH} \sim (r_w^0)^3 / (G r_H^2)$~\cite{Carr:2020gox}.  The black hole relic abundance when the Universe has temperature $T$, expressed as a fraction of the DM relic abundance at $T$, is then
\begin{align}
    f \equiv \frac{\Omega_\text{PBH}(T)}{\Omega_\text{DM}(T)} =&\,
    \frac{n_\text{PBH}(T_n)\,m_\text{BH}}{\Omega_\text{DM}(T_0)\,\rho_\text{crit}(T_0)}
    \frac{g_{\star s}(T_0) T_0^3}{g_{\star s}(T_n) T_n^3}\notag\\
    \approx&\, 
    3.7\times 10^{9} \, \mathfrak{p}  \left(\frac{ T_n}{\text{GeV}}\right)
    \,,
    \label{eq:f}
\end{align}
where the initial number density (at the nucleation temperature of the phase transition, $T_n$) is given by the number of black holes formed per Hubble volume, $n_\text{PBH}(T_n) = 3 \, \mathfrak{p} H^3 / 4\pi$, $\mathfrak{p}$ is the probability that a black hole will form in a given Hubble volume, $T_0$ is the temperature today, $g_{\star s}$ is the effective number of degrees of freedom in entropy, and we assume that the black hole mass does not significantly change between formation and today. In other words, we neglect accretion and Hawking evaporation.
The expression after the second equal sign in \cref{eq:f} does not depend on $T$ any more because the dark matter abundance, $\Omega_\text{DM}$, and the PBH abundance, $\Omega_\text{PBH}$, depend on temperature in the same way.
The parameter $\mathfrak{p}$ depends on the details of the phase transition and can be approximated as the probability of \emph{not} nucleating a true-vacuum bubble in a shrinking false-vacuum pocket. (If new regions with $\ev{\phi} \neq 0$ form before collapse is complete, the accumulating particles will be split into separate populations, each of which is too small to form a black hole.) Quantitatively, $\mathfrak{p} \sim \exp\big[ -\int_0^t \! \upd t' \Gamma_\text{nuc}(t') \big]$. Here, $\Gamma_\text{nu}(t)$ is the nucleation probability, which is typically an exponentially growing function of $\beta t$, where $\beta$ is the parameter describing how fast the bounce action evolves with temperature \cite{Megevand:2017vtb}.  This shows that phase transitions with small $\beta$ -- such as supercooled transitions -- are particularly conducive to black hole formation~\cite{Megevand:2017vtb, DelleRose:2019pgi}.  We discuss this point further in ref.~\cite{Baker:2021sno} and find that $\beta/H \lesssim 4$. 

Taking the conditions on the phase transition together, we see that a black hole can form if (1) the initial radius of the false-vacuum pocket is slightly larger than the Hubble radius, (2) the interior is not vacuum energy dominated, (3) the bubble wall velocity is not too fast ($\gamma_w \lesssim y_\chi \ev{\phi} / T_n$), (4) the phase transition is strong enough to overcome the pressure of the matter trapped in the false vacuum, and (5) the phase transition proceeds slowly ($\beta/H \lesssim 4$).  We note that these conditions are typically not simultaneously satisfied in phase transitions discussed in the literature.  In particular, slow phase transitions (5) typically do not have a small latent heat (2), e.g.\ Ref.~\cite{DelleRose:2019pgi}.  Conversely, phase transitions with a small latent heat (2) do not typically proceed slowly (5), e.g., Ref.~\cite{Megevand:2017vtb}.

However, it is reasonable to think that these conditions could be simultaneously achieved. Additional friction forces acting on the wall (possibly due to further particles present in a complete model) would help accommodate condition (3)~\cite{Baker:2021sno}. Conditions (2) and (4) can both be satisfied if the potential difference $\Delta V$ between the true and false vacua lies in a narrow range~\cite{Baker:2021sno}. Satisfying condition (5), a slow transition, then imposes constraints on the height and width of the barrier between the two vacua.  Finally, for small $\beta$, condition (1) will be satisfied in some regions of the Universe on a statistical basis~\cite{Baker:2021sno}.

It is plausible that even polynomial potentials could satisfy all these conditions simultaneously.  In the limit of small latent heat release $\Delta V$ relative to $T_n^4$, a 4th-order polynomial at $T=0$ can be written as
\begin{align}
    V(\phi) \approx 16 \frac{(v-\phi)^2 \phi^2}{v^4}B + \frac{\phi^2(5v^2 - 14 v \phi + 8\phi^2)}{v^4}\Delta V
    \,,
\end{align}
where $B$ is the barrier height and $v$ is the vev.  In this form we see that we can independently fix the latent heat release, addressing (2) and (4), the barrier height at $T \sim 0$, addressing (5), and the vev. Condition (3) could then be addressed by additional physics that introduces more sources of friction and condition (1) would determine the resulting PBH abundance. A detailed investigation of explicit models that address all conditions is left for further work.

\begin{figure*}
    \centering
    \includegraphics[width=\textwidth]{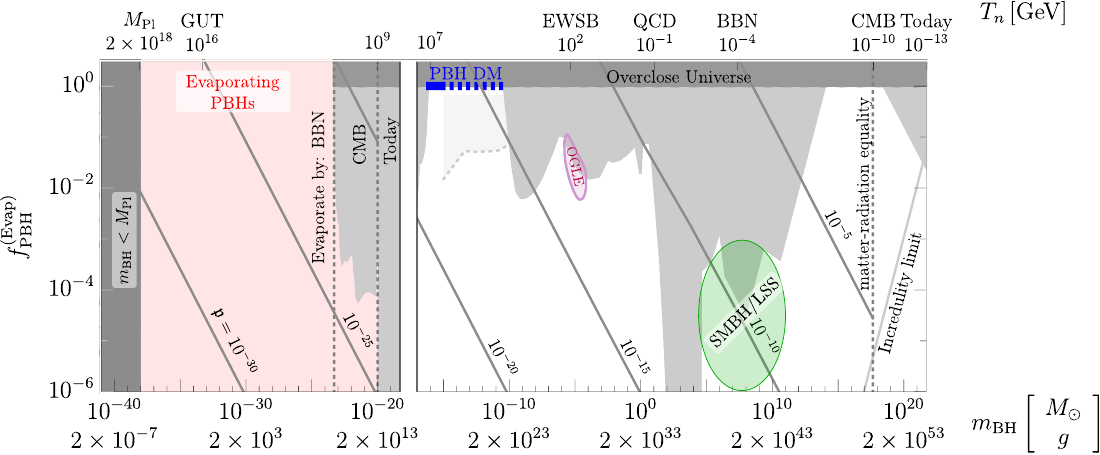}
    \caption{The black hole mass and abundance, $f = \Omega_\text{PBH} / \Omega_\text{DM}$, for black holes produced during a phase transition at temperature $T_n$, and assuming that the probability to form a PBH in a given Hubble volume is $\mathfrak{p}$ (diagonal contours).  In the red region PBHs have already evaporated (and their abundance is shown just before the epoch of evaporation), in the blue region they could account for all the DM in the Universe, in the purple region they could explain the OGLE hint \cite{Mr_z_2017,Niikura:2019kqi}, and in the green region they could provide seeds for supermassive black holes and large scale structure \cite{Bean:2002kx, Hoyle:1966,Ryan:1972,Carr:1984,Afshordi:2003zb,Carr:2020gox}.  The grey region shows the relevant observational constraints~\cite{Carr:2020gox}.}
    \label{fig:mass-density}
\end{figure*}

The dark grey contours in \cref{fig:mass-density} show the PBH mass fraction we predict as a function of the black hole mass $m_\text{BH}$ for different values of $\mathfrak{p}$.  The top axis shows the nucleation temperature $T_n$ corresponding to each black hole mass, assuming that $r_w^0 \sim r_H$.  We see that phase transitions occurring between the Planck scale and the time of matter--radiation equality can form PBHs with masses from the Planck scale up to $10^{18}M_\odot$. Of particular interest may be the following mass ranges: (1) $m_\text{BH} \sim 10^{-15}M_\odot$, where PBHs could account for all the DM in the Universe; (2) below about $10^{-23}M_\odot$, where their evaporation before Big Bang Nucleosynthesis (BBN) would provide an interesting production mechanism for particles (including possibly DM) that interact with the SM only gravitationally; (3) around $10^{10}{M_\odot}$, where PBHs could seed supermassive black holes~\cite{Bean:2002kx} and/or large scale structure formation (LSS)~\cite{Hoyle:1966,Ryan:1972,Carr:1984,Afshordi:2003zb,Carr:2020gox}; (4) around $10^{-5}M_\odot$, where they could explain the possible hint seen by the OGLE survey~\cite{Mr_z_2017,Niikura:2019kqi}.

\textbf{Acknowledgements.}
%
MJB would like to acknowledge support from the Australian Government through the Australian Research Council.
The work of MB, JK, and LM has been partly funded by the German Research Foundation (DFG) in the framework of the PRISMA+ Cluster of Excellence and by the European Research Council (ERC) under the European Union's Horizon 2020 research and innovation programme (grant agreement No.\ 637506, ``$\nu$Directions''). MB and JK have moreover profited from DFG Grant No.\ KO-4820/4-1.

\textbf{Note Added. }
%
In the final stages of this work, ref.~\cite{Gross:2021qgx} appeared showing that dark quark nuggets formed during a confining phase transition may collapse to form dark dwarfs or black holes.

\bibliographystyle{JHEP}
\bibliography{refs}

\end{document}